# Approximate input physics for stellar modelling

Onno R. Pols,* Christopher A. Tout, Peter P. Eggleton and Zhanwen Han
*Institute of Astronomy, Madingley Road, Cambridge CB3 0HA*

**ABSTRACT**
We present a simple and efficient, yet reasonably accurate, equation of state, which at the moderately low temperatures and high densities found in the interiors of stars less massive than the Sun is substantially more accurate than its predecessor by Eggleton, Faulkner & Flannery. Along with the most recently available values in tabular form of opacities, neutrino loss rates, and nuclear reaction rates for a selection of the most important reactions, this provides a convenient package of input physics for stellar modelling. We briefly discuss a few results obtained with the updated stellar evolution code.

## 1 INTRODUCTION

The equation of state (EOS) is one of the most important pieces of physics required as input for a stellar evolution code. Over a substantial region of the $\rho, T$ plane the EOS is relatively simple, and can be computed with an economical code (Eggleton, Faulkner & Flannery 1973; hereinafter EFF). But along the boundary of this region, at fairly low temperature and high density, there are three important processes which were either ignored or treated very crudely in EFF: dissociation of molecular hydrogen, Coulomb interactions, and pressure ionization. A simple contribution for the $H_2$ molecule has in fact been included in the code for several years, though not published. Coulomb interactions between the charged particles provide the major non-ideal correction to the pressure at the densities and temperatures encountered in stars of around a solar mass or less, while they also crucially influence the properties of matter at high densities and low temperatures.

In this paper (Section 2) we give a simple but adequate approximation to the Coulomb interactions, and we describe a somewhat improved, though still very approximate, treatment of pressure ionization. The result is an EOS which, like that of EFF, is completely explicit in its input variables (temperature and electron degeneracy), and which is thermodynamically consistent, and very easy to compute. It may be reasonably accurate for lower-mass stars ($\sim 0.1 - 1 \, M_\odot$), as well as for the initial phases of cooling white dwarfs, both of which were not very accurately treated by the formulation of EFF. It appears to agree reasonably well with the much more detailed but time-consuming calculations of Rogers & Iglesias (1992), Iglesias, Rogers & Wilson (1992) and Rogers (1994) (hereinafter collectively OPAL), and of Mihalas, Däppen & Hummer (1988) and Däppen et al. (1988) (hereinafter collectively MHD).

We have incorporated the improved EOS into the evolution program of Eggleton (1971, 1972, 1973). This program has been modified many times in over 20 years, though the details of the modifications have not normally been published except occasionally and briefly in relation to specific stellar evolutionary problems. Considerable care was taken during these modifications to keep the code simple: in fact, most changes consisted of *removing* bits of code that were necessary when, for example, double precision was an expensive luxury, or 200 Kbytes was the maximum space available. A recent paper (Han, Podsiadlowski & Eggleton 1994, hereinafter HPE) summarizes the main developments in the code, and so we will not repeat them here. However, since HPE we have improved not only the EOS, but also the opacity tables, the neutrino loss rates and the nuclear reaction network, and so we describe these changes here, in Section 3.

In Section 4 we apply the resulting evolution code to construct a zero-age main sequence (ZAMS), and evolutionary sequences for a wide range of masses. We discuss these results only briefly; we hope to consider various aspects of them in a later paper.

## 2 THE EQUATION OF STATE

The main reason for the simplicity of the code of EFF was that a Helmholtz free energy $F$ was used in which only the free-electron contribution was at all complicated. All atomic species, ionized or neutral, contributed only simple Maxwell–Boltzmann terms. The electrons alone were allowed a certain degree of complexity, both as Fermi–Dirac particles and as the agents of pressure ionization. This meant first that it was very convenient to use as independent variables the temperature $T$, the electron degeneracy $\psi$ (the Gibbs free energy divided by $kT$), and the abundances (neutral plus ionized, and so independent of the previous two variables); and secondly that the ionization equilibrium could be calculated *explicitly*, since the equilibrium of, say, H$^+$ + e $\leftrightarrow$ H is not dependent (as, physically, it ought to be) on the concentration of, say, He, He$^+$ or He$^{++}$. Thus it is not necessary to solve simultaneously a complicated and highly non-linear set of equations for the ionization equilibrium of all species. The inclusion of H$_2$, still as Maxwell–Boltzmann particles, only complicates this slightly, requiring one to solve a quadratic rather than a linear equation to find the overall fraction of electrons that are free.

* E-mail: onno@ast.cam.ac.uk

The quantities that a stellar evolution algorithm requires to be calculated from the EOS are only pressure $p$, density $\rho$, adiabatic gradient $\nabla_{\rm a}$, and specific heat at constant pressure $C_p$, and also (in some algorithms) their derivatives with respect to the independent variables. For some purposes, such as the study of oscillations, one may also require the compressibility $\gamma$ and its derivatives. Of course, first derivatives of $\nabla_{\rm a}$, $C_p$ and $\gamma$ are in effect third derivatives of $F$, and so we see it as important for thermodynamic consistency that we start with an expression for $F$ which is continuous and sufficiently simple analytically that even third derivatives can be extracted without great labour.

## 2.1  Algorithm

The formulation of our EOS, like that of MHD, is based on the principle of Helmholtz free energy minimization (cf. Graboske, Harwood & Rogers 1969, hereinafter GHR). We start from a free energy of the form

$$F(V, T, N_{\rm e}, N_i) = -\frac{1}{3} a T^4 V$$
$$+ kT \sum_i N_i \left( \ln \frac{N_i h^3}{\omega_i V (2\pi m_i kT)^{3/2}} - 1 + \frac{\chi_i}{kT} \right)$$
$$+ F_{\rm e}(N_{\rm e}, V, T) + \Delta F_{\rm e}(N_{\rm e}, V, T). \quad (1)$$

Here, $V$ is the volume per unit mass (the reciprocal of the density $\rho$). Other variables and constants have their usual meanings, unless specified otherwise later on in this section. The first term on the right is due to radiation, and is relatively trivial to include. The second term is the contribution from atomic and molecular particles. The $N_i$ are the numbers of particles per unit mass of species $i$. The only non-trivial species that we include are H$^+$, H, H$_2$, and He$^{++}$, He$^+$, He. We also include seven heavier elements, C, N, O, Ne, Mg, Si and Fe, which are assumed to be fully ionized, at all temperatures and densities. Each of these elements contributes one free electron for every two baryons, except of course for Fe which contributes fractionally more. It would not in fact be conceptually more difficult (although certainly more laborious) to include the ionization of more nuclear species. But the effect on the EOS of the $\lesssim 2$ per cent of material that is 'metals' is slight. Of course, the effect of the state of ionization of the metals on the *opacity* is enormous, but we take opacities in tabular form from work that treats the EOS much more rigorously (Section 3).

The energies $\chi_i$ of ionization or dissociation for the six non-trivial species are respectively (in eV) 0, $-13.60$, $-31.68$, 0, $-54.40$, $-78.98$ (the zero-point of energy taken, for convenience, to be the fully ionized state). The partition functions $\omega_i$ are in general functions of temperature, but except for H$_2$ we approximate these by the statistical weights of the ground states: $\omega_i = 2$ for H and He$^+$ and $\omega_i = 1$ for the other species. For H$_2$ it is necessary to take into account the low-energy rotational and vibrational modes; we therefore adopt a partition function which derives from Vardya's (1960) expression for the pressure equilibrium constant $K_p({\rm H}_2)$:

$$\omega_{{\rm H}_2} = 6608.8 \, \zeta(T) \left( \frac{kT}{D_{{\rm H}_2}} \right)^{5/2} e^{D_1/kT - (D_2/kT)^2 + (D_3/kT)^3}, (2)$$

where $D_{{\rm H}_2} = 2\chi_{\rm H} - \chi_{{\rm H}_2} = 4.48$ eV, the binding energy of molecular hydrogen relative to atomic hydrogen, and the values of the constants $D_1, D_2, D_3$ are (in eV) 0.448, 0.1562, 0.0851. The factor

$$\zeta(T) = 1 - \left( 1 + \frac{D_{{\rm H}_2}}{kT} \right) e^{-D_{{\rm H}_2}/kT}, \quad (3)$$

taken from Webbink (1975), truncates the partition function which would otherwise diverge for large $T$.

The third term of equation (1), $F_{\rm e}$, is the term that treats electrons as pure Fermi–Dirac (FD) particles:

$$F_{\rm e} = N_{\rm e} \psi kT - V p_{\rm e}. \quad (4)$$

$N_{\rm e}$ is the number of free electrons per unit mass. Both $n_{\rm e}$ ($\equiv N_{\rm e}/V$) and $p_{\rm e}$, the electron pressure, are the normal FD integrals (see Chandrasekhar 1939), which are explicit functions of $\psi, T$. We approximate these in the same way as EFF. For *analytical* purposes, the right-hand side of equation (4) is to be seen as a function of the independent variables $N_{\rm e}, V, T$. For *computational* purposes, however, it is much more convenient to take $\psi, T$ as independent variables. $N_{\rm e}$ has then to be computed from ionization equilibrium (see below), so that subsequently $V$, or equivalently $\rho$, is determined from the FD integral for $n_{\rm e}$. EFF gave simple but accurate approximations for the three FD integrals, $n_{\rm e}, p_{\rm e}$, and a third one for the internal energy of the free electrons. However, since EFF we have found that greater numerical accuracy can be achieved if we evaluate a different integral, related to $s_{\rm e} \equiv S_{\rm e}/V$, where $S_{\rm e}$ is the entropy per unit mass of the free electrons (see Appendix A). It should be noted that the identities

$$\left( \frac{\partial p_{\rm e}}{\partial \psi} \right)_T = n_{\rm e} kT, \quad \left( \frac{\partial p_{\rm e}}{\partial T} \right)_\psi = s_{\rm e} + n_{\rm e} \psi k \quad (5)$$

are satisfied exactly by these approximations.

The fourth term of equation (1), $\Delta F_{\rm e}$, is the term into which we endeavour to put all the non-ideal modifications which will make the EOS accurate enough for stellar interior purposes. Because it involves only $N_{\rm e}$ and not the $N_i$, it is guaranteed to be simple to use: the degrees of ionization and of dissociation will be explicit functions of $\psi$ and $T$. The question is whether we can find a functional form that will give sufficient accuracy. We address this in the next subsection.

From $F$ are obtained the pressure, entropy and chemical potentials, according to the usual rules:

$$p = -\left( \frac{\partial F}{\partial V} \right)_{T, N_{\rm e}, N_i}$$

$$= \tfrac{1}{3}aT^4 + kT\frac{\Sigma_i N_i}{V} + p_e - \left(\frac{\partial \Delta F_e}{\partial V}\right)_{T,N_e}, \qquad (6)$$

$$\begin{aligned} S &= -\left(\frac{\partial F}{\partial T}\right)_{V,N_e,N_i} \\ &= k\sum_i N_i \left[\frac{5}{2} - \ln\frac{N_i h^3}{\omega_i V(2\pi m_i kT)^{3/2}} + \frac{\partial \ln \omega_i}{\partial \ln T}\right] \\ &\quad + \tfrac{4}{3}aT^3 V + S_e - \left(\frac{\partial \Delta F_e}{\partial T}\right)_{V,N_e}, \end{aligned} \qquad (7)$$

$$\begin{aligned} \mu_i &= \frac{1}{kT}\left(\frac{\partial F}{\partial N_i}\right)_{T,V,N_e} \\ &= \ln\frac{N_i h^3}{\omega_i V(2\pi m_i kT)^{3/2}} + \frac{\chi_i}{kT}, \end{aligned} \qquad (8)$$

$$\begin{aligned} \mu_e &= \frac{1}{kT}\left(\frac{\partial F}{\partial N_e}\right)_{T,V,N_i} \\ &= \psi + \Delta\mu_e; \quad \Delta\mu_e \equiv \frac{1}{kT}\left(\frac{\partial \Delta F_e}{\partial N_e}\right)_{T,V}. \end{aligned} \qquad (9)$$

To obtain the electron terms that derive from $F_e$, we have used the identities (5). The terms $\partial \ln \omega_i / \partial \ln T$ in equation (7) are zero except for $H_2$.

The degrees of ionization and of molecular dissociation are obtained by minimizing the free energy subject to the conservation constraints

$$2N_{H_2} + N_H + N_{H+} = X/m_H, \qquad (10)$$

$$N_{He} + N_{He+} + N_{He++} = Y/4m_H, \qquad (11)$$

and

$$N_e = N_{H+} + N_{He+} + 2N_{He++} + Z/2m_H, \qquad (12)$$

where $X, Y, Z$ (summing to unity) are the usual abundance fractions by mass of hydrogen, helium and 'metals'. These lead to the conditions, or 'Saha equations':

$$2\mu_H = \mu_{H_2},$$

$$\text{i.e. } \frac{N_H^2}{N_{H_2} N_e} = \frac{4}{\omega_{H_2}}\left(\frac{\pi m_H kT}{h^2}\right)^{\frac{3}{2}}\frac{1}{n_e}e^{-4.48/kT}, \qquad (13)$$

$$\mu_H = \mu_e + \mu_{H+}, \text{ i.e. } \frac{N_{H+}}{N_H} = \frac{1}{2}e^{-\psi-\Delta\mu_e-13.60/kT}, \qquad (14)$$

$$\mu_{He} = \mu_e + \mu_{He+}, \text{ i.e. } \frac{N_{He+}}{N_{He}} = 2e^{-\psi-\Delta\mu_e-24.58/kT}, \qquad (15)$$

and

$$\mu_{He+} = \mu_e + \mu_{He++}, \text{ i.e. } \frac{N_{He++}}{N_{He+}} = \frac{1}{2}e^{-\psi-\Delta\mu_e-54.40/kT}. \qquad (16)$$

In equations (13)–(16) and subsequently, we give the $\Delta\chi_i$ in eV; thus $k$ in some places should be in eV K$^{-1}$ rather than erg K$^{-1}$. In equation (13), we have divided both sides by a factor $N_e$, in order that the left-hand sides of all the equations (13)–(16) should be explicit functions of $\psi, T$ only, given that $\Delta\mu_e$ also has such a form (see below). Thus equations (10)–(16) are seven simultaneous equations for the seven unknown abundances. They have a particularly simple form: eliminating six of them in favour of, say, $N_H$, we have only a quadratic equation in $N_H$ to solve. All thermodynamic properties can then be derived using equations (6) and (7), and the internal energy follows from $U = F + TS$.

## 2.2 Non-ideal corrections

The fourth term in equation (1), $\Delta F_e$, contains the contribution to the EOS owing to the non-ideal effects of Coulomb interactions and pressure ionization. As was demonstrated above, the simplicity and explicit nature of the algorithm rely on the chemical potential deriving from this term being dependent only on $\psi$ (through the electron density $n_e$) and $T$. We therefore seek a function $g(n_e, T)$ in order to approximate both these effects with a $\Delta F_e$ of the form

$$\Delta F_e = -N_e kT\, g(n_e, T), \qquad (17)$$

which, using equation (9), implies that

$$\Delta\mu_e = -g - n_e \frac{\partial g}{\partial n_e}. \qquad (18)$$

### 2.2.1 Coulomb interactions

Formulations for Coulomb interactions have been developed in the limits of weak and strong interaction. In the weak limit (cf. MHD; GHR), one usually defines the plasma interaction parameter

$$\Lambda = \frac{\zeta^2 e^2}{r_s kT}; \quad \zeta^2 = \frac{\sum_i N_i z_i^2 + N_e \theta_e}{\sum_i N_i}, \qquad (19)$$

where $r_s$ is a generalized screening length: $r_s = (4\pi\zeta^2 e^2 \sum_i N_i/kTV)^{-1/2}$. Here, and in the rest of this section, the summation $i$ extends over all *ions*, with charges $z_i$. The factor $\theta_e = \partial \ln n_e/\partial \psi$ corrects for electron degeneracy (it equals 1 in the limit $\psi \to -\infty$, and 0 in the limit $\psi \to \infty$). In this limit, the free energy contribution is

$$\Delta F_C = -\sum_i N_i\, kT\, \frac{\Lambda}{3}\cdot\tau, \qquad (20)$$

where the factor $\tau \leq 1$, defined by GHR, takes into account the finite ion size. This approximation is valid for $\Lambda \lesssim 0.3$, and reduces to the Debye–Hückel approximation (cf. Cox & Giuli 1968) for $\Lambda \ll 1$ and $\psi \ll 0$.

In the strong interaction limit, for $\psi \gg 0$, it is customary to define a different interaction parameter for a one-component plasma,

$$\Gamma_i = \frac{z_i^2 e^2}{r_i kT}, \qquad (21)$$

where $r_i$ is the ion-sphere radius $r_i = (4\pi N_i/3V)^{-1/3}$. The free energy is then the sum of the contributions of the different plasma components:

$$\Delta F_C = \sum_i \Delta F_{Ci} = -\sum_i N_i kT g(\Gamma_i). \qquad (22)$$

In the limit $\Gamma_i \to \infty$, corresponding to a zero-temperature lattice structure, $g(\Gamma_i) \approx 0.9\Gamma_i$ (Cox & Giuli 1968). Slattery, Doolen & DeWitt (1980; hereinafter SDD) have performed Monte Carlo calculations of the excess free energy of a one-component plasma in the liquid phase, for $1 < \Gamma_i < 160$, and in the solid phase, for $160 < \Gamma_i < 300$, and given an analytical expression for $g(\Gamma_i)$ as a fit to their results.

Although the free energy term $\Delta F_C$ clearly depends on all ion numbers $N_i$, we can arrive at an expression of the form (17) by making two approximations. First, we replace the ratios $\Sigma_i N_i/N_e$ and $\Sigma_i N_i z_i^2/N_e$ by $N_0/N_{e0}$ and $\Sigma_j(X_j Z_j^2/m_j)/N_{e0}$ respectively, where $N_e = \Sigma_j X_j Z_j/m_j = 0.5(1+X)/m_H$ is the total number of electrons (bound or free) and $N_0 = \Sigma_j X_j/m_j$ is the total number of nuclei (atomic or ionic). The summation here extends over all *nuclear* species $j$ (with charge $Z_j$ and mass $m_j$), so that these quantities are independent of $\psi$ and $T$. Although quantities like $\Sigma_i N_i$ may vary greatly, their ratios will do so much less, so that this approximation should be reasonably accurate even in regions of (partial) de-ionization. Secondly, we write the sum $\Sigma_i N_i g(\Gamma_i)$ as $\Sigma_i N_i \cdot g(\Gamma)$, where $\Gamma$ is an average plasma interaction parameter. The usual way of averaging would be $\Gamma = \Sigma_i N_i \Gamma_i / \Sigma_i N_i$, but we find it more convenient to use $\Gamma = \zeta^2 e^2/r_0 kT = (r_s/r_0)\Lambda = (\Lambda/\sqrt{3})^{2/3}$, where $r_0 = (4\pi \Sigma_i N_i/3V)^{-1/3}$ is the average inter-ion distance.

With these approximations, the explicit expression for $\Gamma$ is

$$\Gamma = \left(\frac{4\pi}{3}\right)^{1/3} \frac{e^2 n_e^{1/3}}{kT} \left(\frac{N_{e0}}{N_0}\right)^{2/3} \left(\frac{\Sigma_j(X_j Z_j^2/m_j)}{N_{e0}} + \theta_e\right), \quad (23)$$

and we model Coulomb interactions with a term in $\Delta F_e$ of the form

$$\Delta F_C = -N_e kT \frac{N_0}{N_{e0}} g_C(\Gamma), \quad (24)$$

$$g_C(\Gamma) = \frac{a_1 \Gamma}{[(\Gamma/\Gamma + a_3)^{1/a_2} + (a_1\sqrt{3/\Gamma})^{1/a_2}]^{a_2}}. \quad (25)$$

This term has the correct asymptotic form in the limits of both large and small values of $\Gamma$. The values of the constants $a_1$, $a_2$ and $a_3$ in $g_C(\Gamma)$ are found by fitting this expression to equation (20) for $0.01 < \Gamma < 0.3$ and to the expression for $g(\Gamma)$ given by SDD for $1 < \Gamma < 160$. The values $a_1 = 0.89752$ (from SDD), $a_2 = 0.768$ and $a_3 = 0.208$ give a correspondence better than 4 per cent over the whole region. The specific heat $C_p$, which involves the second derivative of $g_C$ and is the quantity most sensitive to the Coulomb term, corresponds within 2 per cent to the values derived from SDD. We are therefore confident that this treatment is fairly accurate up to $\Gamma = 160$, which corresponds to the phase transition to the crystalline state.

### 2.2.2 Pressure ionization

In order to model pressure ionization, we need to choose a $\Delta F_e$, and consequential $\Delta \mu_e$ (equation 9), which will oppose the tendency in equation (14) for hydrogen to become steadily more de-ionized as $\psi$ increases at constant $T$. The behaviour we want is that hydrogen should start to ionize again once the density becomes so large that atoms are separated by $\lesssim a_0$, the Bohr radius. We therefore seek a simple term with the properties that,

(a) for $n_e \ll a_0^{-3}$, the term is negligible;
(b) for $n_e \sim a_0^{-3}$, $\Delta\mu_e$ should roughly cancel the standard term $\psi + 13.60/kT$;
(c) for $n_e \gg a_0^{-3}$, the term should become large and negative, but increasing sufficiently slowly with increasing $\psi$ that not just $N_e$ and $n_e$ but also $\rho = n_e/N_e$ increase monotonically.

The last condition is required to avoid an awkward multivaluedness in the EOS. If the pressure-ionization term were allowed to increase the degree of ionization, and hence $N_e$, too rapidly, then $\rho$, after increasing with increasing $\psi$, would decrease temporarily, before increasing again once the hydrogen is fully ionized. Such multivaluedness would imply some kind of phase transition, but this does not appear to happen with the much more detailed EOSs of MHD and OPAL. We cannot claim, however, to have been very successful in fulfilling (c): our best formulation for (a) and (b) still leads to multivaluedness, compressed into a small region of the $\rho, T$ plane, as noted in Section 2.3.

The following term, containing a number of constants ($c_1 \ldots c_4$) which can be varied experimentally to give a good fit, appears to be adequate:

$$\Delta F_{PI} = -N_e kT\, g_{PI}(x,y) \quad (x \equiv n_e m_H, y \equiv 13.60/kT), \quad (26)$$

$$g_{PI}(x,y) = e^{-(c_1/x)^{c_2}} [y + \psi + c_3 \ln(1 + x/c_4)]. \quad (27)$$

The exponential at the beginning of equation (27) takes care of point (a), for suitable $c_1, c_2$. When $x$ is sufficiently large that the exponential is effectively unity, the terms $y$ and $\psi$ take care of point (b), provided that the logarithmic term following is still small. Finally, when $x$ is large enough for the logarithmic term to become important, point (c) is nearly, though not entirely, satisfied for suitable $c_3, c_4$. The choice of values $c_1, c_2, c_3, c_4 = 3, 0.25, 2, 0.03$ ($c_1$ and $c_4$ in units of $\text{g cm}^{-3}$) provides reasonable agreement with the location of ionization zones according to MHD.

The expression (27) was chosen to ensure that the *hydrogen* becomes fully pressure-ionized at high density. However, at somewhat higher temperature and density it also, somewhat crudely, takes care of the *helium*. The formulation might be improved by adding further terms like (27) which use the helium ionization energy (54.40 eV) instead of the hydrogen value (13.60 eV), but such elaboration appears unnecessary. Furthermore, we do not separately treat the pressure dissociation of $H_2$ which should set in at $\rho \gtrsim 10^{-2} \text{ g cm}^{-3}$. Therefore the ratio $N_{H_2}/N_H$ continues to increase towards higher density, but in this region of $\rho, T$ the validity of our EOS becomes questionable anyway. However, at high density where hydrogen is fully ionized, the $H_2$ abundance is also negligible.

When the material is fully ionized at high pressure, the equation of state in effect becomes simple again: we expect to get the same EOS as we would get by simply making the assumption that ionization is total. However, our term (26), (27) affects not only the ionization via $\Delta\mu_e$, but also the pressure and entropy via equations (6) and (7). This contribution does not go to zero, as it should, at full ionization, unless we subtract from the free energy a compensating term thus:

$$\Delta F_e = \Delta F_C + \Delta F_{PI}(N_e, V, T) - \Delta F_{PI}(N_{e0}, V, T), \quad (28)$$

where $N_{e0}$ is the *total* number of electrons, bound or free, as defined above. Since the extra term does not depend explicitly on $N_e$, it does not affect $\Delta\mu_e$, but it ensures that the effect of pressure ionization on both pressure and entropy goes to zero as $N_e \to N_{e0}$. Of course, the Coulomb screening term continues to contribute to both pressure and entropy. Hence equation (28), along with equations (23)–(27), is the extra term which we incorporate in equation (1).

Our use of $\psi$ in equation (27) slightly conflicts with our need to subtract the last term in equation (28), because,

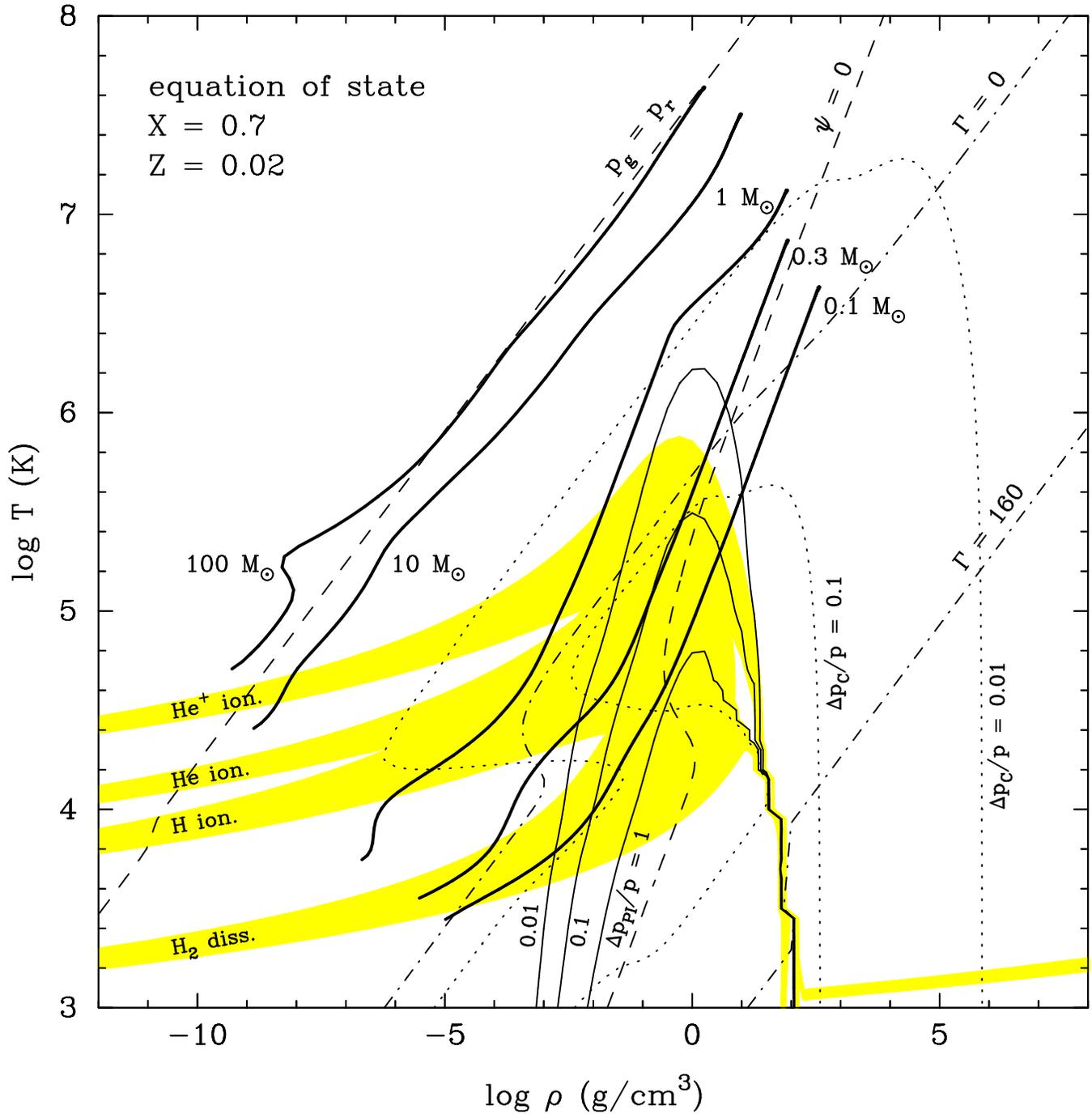

**Figure 1.** The equation of state in the $\log \rho, \log T$ diagram, for an element mixture with $X = 0.7$ and $Z = 0.02$. Dashed lines indicate where radiation pressure dominates over gas pressure (to the left of the line $p_g = p_r$), and where electron degeneracy is important (to the right of the line $\psi = 0$). Dash-dotted lines show where the plasma-interaction parameter $\Gamma$ (equation 23) equals 1 and 160. The dotted lines indicate regions where the pressure term due to Coulomb interactions contributes to the ideal pressure (see text) by more than 0.01 and 0.1 fractionally. The thin solid lines similarly indicate the regions where the pressure term due to pressure ionization contributes to the ideal pressure by more than 0.01, 0.1, and 1. Shaded areas show regions where the ratios $N_{H_2}/N_H$, $N_H/N_{H^+}$, $N_{He}/N_{He^+}$, and $N_{He^+}/N_{He^{++}}$ lie in the range 0.1 to 10. Shown as thick solid lines are a few ZAMS models of stellar interiors of different masses, calculated with the updated stellar evolution code (Section 4).

whereas we have (in fact, as an input variable) the $\psi$ that is consistent, via the FD integral, with the quantities $N_e, V, T$ of the second-last term in equation (28), we can only obtain the corresponding $\psi_0$ for the last term if we invert the corresponding FD integral for $n_{e0}$ as a function of $\psi_0, T$. This would be laborious, and not really necessary since equation (27) is very ad hoc. Consequently, we use (but only in this section of the code) a much weaker approximation to the FD integral for both $\psi(n_e, T)$ and $\psi_0(n_{e0}, T)$:

$$\psi = 2\sqrt{f}; \quad f = 1.07654\, xy^{3/2}(1 + 0.61315\, xy^{3/2})^{1/3}. \quad (29)$$

This expression approximates $\psi$ in the region of large degeneracy, where it is important in the above context.

### 2.3 Selected results and discussion

Fig. 1 summarizes the effects of the various contributions to the EOS. Shown as a function of $\rho, T$ are the regions where radiation pressure, electron degeneracy and Coulomb interactions influence the EOS. The dotted lines show the regions where Coulomb interactions contribute to the ideal pressure (i.e. the pressure obtained without the term $\Delta F_e$) by more than 1 and 10 per cent. This contribution is always negative, but never leads to a negative pressure. Note that, because of degenerate electron pressure, the relative pressure contribution decreases again for large enough density, although it continues to increase in absolute value. However, for $\Gamma \gtrsim 1$ the *thermal* properties of the gas become increasingly dominated by the Coulomb term – in particular $C_p$ increases by more than a factor of 2 as $\Gamma$ goes from 1 to 160. For still larger values of $\Gamma$, crystallization sets in, while for temperatures $kT \lesssim \hbar\omega_p$, where $\omega_p$ is the plasma frequency, a quantum-mechanical treatment of lattice vibrations is necessary, and our EOS becomes unreliable. A simple model accounting for these effects would in fact not be difficult to implement, and is potentially interesting for the study of cooling white dwarfs, but this might also require a better treatment of pressure ionization in the envelopes. Note that, for a realistic C–O mixture, the line $\Gamma = 160$ lies at $\Delta \log T \approx 1.2$ higher than for the H–He mixture shown here.

The shaded areas in Fig. 1 are regions of partial ionization and dissociation, namely regions where $N_{H_2}/N_H$, $N_H/N_{H+}$, $N_{He}/N_{He+}$, and $N_{He+}/N_{He++}$ lie in the range 0.1 to 10. The thin solid lines show regions where our treatment of pressure ionization modifies the ideal pressure by more than 1, 10 and 100 per cent (always in a positive sense). Although any prescription for pressure ionization necessarily leads to pressure and entropy contributions, a comparison with the OPAL EOS (see below) suggests that, in our case, they are largely spurious. Pressure ionization occurs at roughly the same density for all species, whereas, in the much more detailed EOS of MHD, $He^+$, for instance, dissociates at higher density than H. For $\log T \lesssim 4.5$, pressure ionization sets in so abruptly that a density inversion cannot be avoided. This leads to a discontinuity in the $\rho, T$ plane, shown as the very thick solid line in Fig. 1. Around this region, and down to densities $\gtrsim 10^{-2}\,\mathrm{g\,cm^{-3}}$, the validity of our EOS is very dubious, although it remains thermodynamically consistent everywhere. As a curious aside, the failure of our algorithm to pressure-dissociate $H_2$ partly compensates for the spurious pressure term in this region!

In Figs 2 and 3 we compare values that we obtain from the present code with values from the OPAL EOS (made available through anonymous ftp by F. Rogers). Fig. 2 shows contours of the difference between our pressure and that of OPAL, as a fraction of our pressure. It can be seen that the agreement is better than 2 per cent over most of the region where data are available, although our pressure seems to be systematically too large. Part of this difference may be attributed to the assumption that all 'metals' are fully ionized, although this error can be at most of order $Z/2$, where $Z$ denotes metallicity, or 1 per cent for the assumed solar mixture. The largest deviation (up to 5 per cent) seems to occur in the region where our pressure ionization term contributes most strongly to the pressure. Fig. 3 similarly shows differences between the $\nabla_a$ of our EOS and that of OPAL. These values also agree to better than 1 per cent over most of the region, although a systematic deviation of order 0.04 (about 10 per cent) occurs at low temperatures, just below the hydrogen ionization band. It appears to be due to the fact that $N_H/N_{H+}$ increases too rapidly towards lower $T$, or larger $\rho$. This may again be (partly) attributed to the assumption of fully ionized metals: supposing that $N_H/N_{H+}$ is correct as a function of $n_e, T$, then, if $N_e$ is larger than it ought to be, $\rho = n_e/N_e$ is correspondingly too small, and hence $N_H/N_{H+}$ will be too large at a fixed $\rho$.

## 3 OPACITIES, NEUTRINO LOSSES, AND NUCLEAR REACTION NETWORK

Entirely separately from the EOS, we incorporate tables of radiative opacity which derive from OPAL, and from Alexander (1994, private communication; see also Alexander & Ferguson 1994a, 1994b) for low temperatures where molecular opacity is important. Their values overlap for $\log T$ between 3.8 and 4.1, and agree to within a few per cent in this region. We adopt the OPAL values at $\log T = 4.1$ and the Alexander values at $\log T = 3.8$, and at intermediate temperatures we take values weighted linearly in $\log T$. At high temperatures and low densities that fall outside the boundaries of these tables, we assume simple electron scattering (corrected for relativistic effects). Degenerate electron conductivities are taken from Itoh et al. (1983), supplemented by values from Hubbard & Lampe (1969) in the partially degenerate region. From the radiative and conductive opacities we construct (by reciprocal addition) a rectangular table in $\log \rho, \log T$ space, the first incrementing in steps of 0.25 from $-12$ to 10, and the second in steps of 0.05 from 3.0 to 9.3. We use quadratic interpolation in the OPAL data to achieve this. Outside the subset of this area for which values are available, our values are obtained by extrapolation; such values of course have no validity, but we believe that a rectangular, equal-interval mesh is much the most convenient format for a stellar evolution code. Our mesh is sufficiently fine that we can, during an evolution run, use only linear interpolation for a sufficiently accurate model. Of course, for such purposes as stability analysis, one needs a more sophisticated approach in order to obtain derivatives that are reliable. The code does not note whether some point in the star has evolved into the invalid extrapolated area, but, since the valid area does in practice cover most of the region where stellar interiors are expected to lie, this is not a big problem.

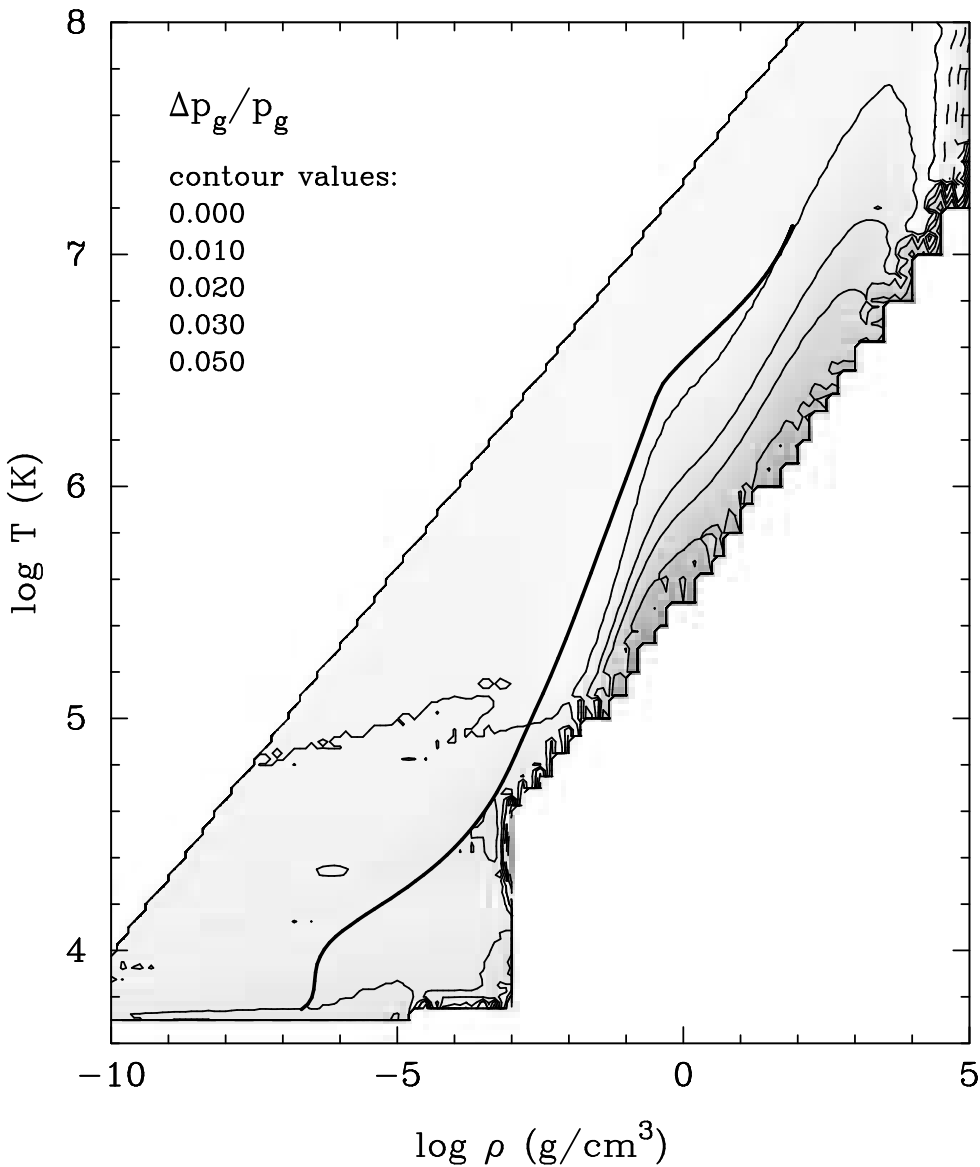

**Figure 2.** The fractional difference $(p_g - p_{g,\text{OPAL}})/p_g$ between the gas pressure obtained with our EOS and that of OPAL, as a function of $\log \rho, \log T$. The grey-scale is proportional to this quantity, and contours are drawn at the indicated absolute values; solid lines are positive or zero values and dashed lines are negative values. The composition is as in Fig. 1. The thick solid line is a 1-$M_\odot$ ZAMS model. The crenellated boundary towards the lower right is the limit of values supplied by OPAL.

It is in the deep interiors of low-mass main-sequence stars that one is most likely to pick up values of opacity outside the valid area. Somewhat fortuitously, at the lowest masses such interiors are entirely convective, and thus the interior opacity (but not the surface opacity) is largely irrelevant.

Opacity values are now available for a considerable range of metallicity $Z$, as well as at rather smaller intervals of hydrogen abundance $X$ than heretofore. In any one evolutionary run it is normally only necessary to have an opacity table for a single zero-age metallicity $Z_0$. For each $Z_0$, we use a single abundance variable to interpolate in composition; this variable is $X$ in regions where $X > 0$ and $Y$ otherwise. We make use of the fact that, during helium burning, there exists a fairly unique relationship between the abundances of He, C and O, more or less independent of stellar mass.

For each $Z_0$ we have tables for $X = 0.70, 0.35, 0.10, 0.03$ and 0, and for $Y = 0.5 - Z_0$, $0.2 - Z_0$ and 0, and we interpolate linearly in $X$ or $Y$.

Neutrino loss rates are taken from a series of papers by N. Itoh and collaborators. Rates for the photo- and pair-neutrino processes are taken from Itoh et al. (1989) and for the plasma process from Itoh et al. (1992). These rates are all independent of the chemical composition. Rates for $\nu$-pair bremsstrahlung are taken from Itoh & Kohyama (1983) for the degenerate liquid region, and from Munakata, Kohyama & Itoh (1987) for the partially degenerate region. The bremsstrahlung rates are composition-dependent, but for the partially degenerate region this dependence is a simple proportionality to $\langle Z^2/A \rangle$. We therefore find it most convenient to construct two tables, one for the rate $\epsilon_\text{p}$ of photo-,

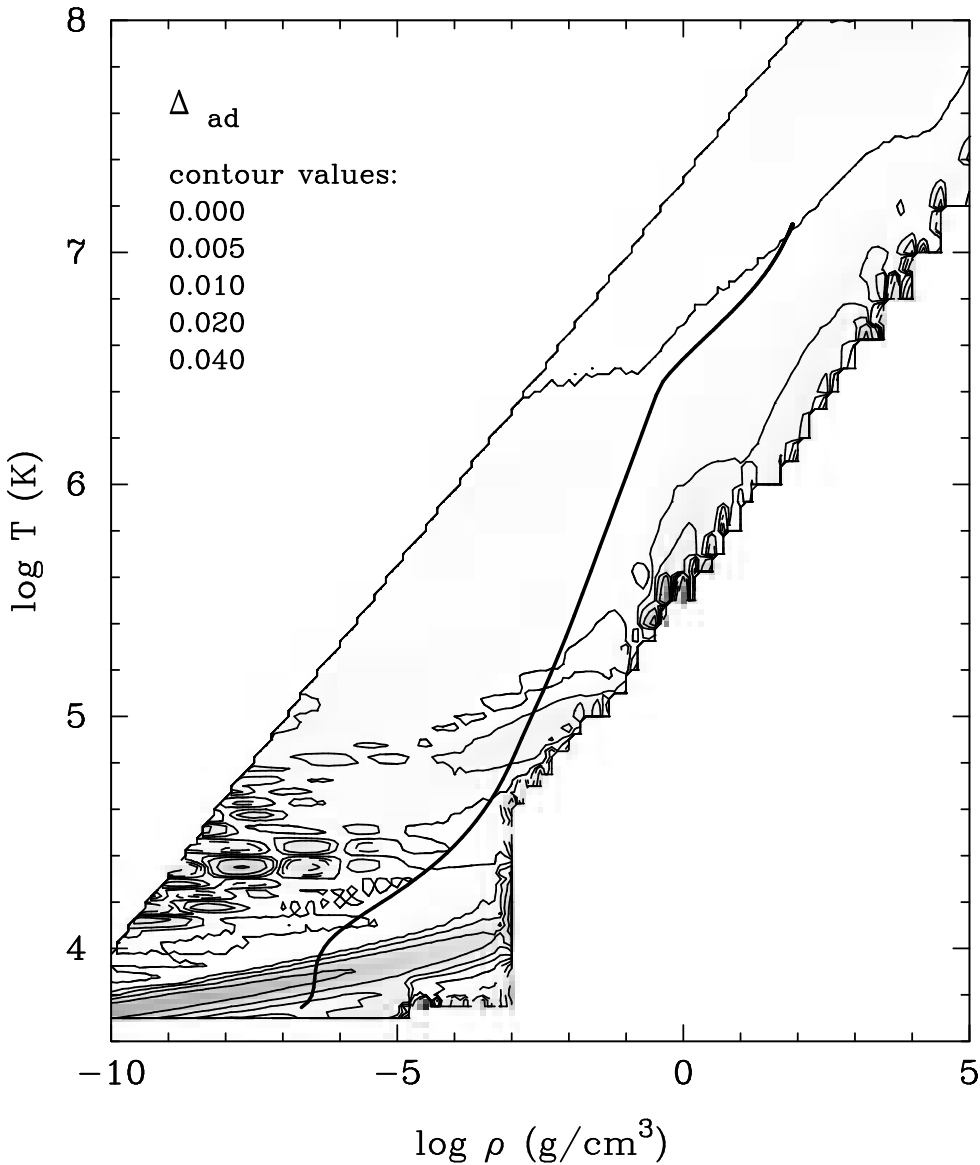

**Figure 3.** The absolute difference $\nabla_a - \nabla_{a,\mathrm{OPAL}}$ between the adiabatic temperature gradient obtained with our EOS and that of OPAL, as a function of $\log \rho, \log T$. Only the gas component is taken into account. Legend as for Fig. 2.

pair- and plasma processes combined, and one for the rate $\epsilon_b$ of bremsstrahlung neutrinos for a single element, which we take to be oxygen. The total neutrino loss rate is then taken to be $\epsilon_\nu = \epsilon_p + \epsilon_b \langle Z^2/A \rangle / 4$, since $Z^2/A = 4$ for $^{16}$O. The value thus obtained only deviates from the correct rate at very high density and relatively low temperature, and when the 'mean' composition is very different from oxygen. Such densities and temperatures are only encountered in the cores of very evolved stars which are predominantly composed of oxygen, so the error introduced is usually small. The tables for $\epsilon_p$ and $\epsilon_b$ are constructed on the same grid as for opacities, for $7 < \log T < 10$ and $0 < \log \rho < 10$.

It is a feature of the Eggleton code (HPE) that the composition equations are solved *simultaneously* with the structure (and also simultaneously with the mesh-spacing). Even with modern computing power it would be uneconomical to follow a large number of nuclear species in this way. We therefore limit ourselves to five major species – $^1$H, $^4$He, $^{12}$C, $^{16}$O and $^{20}$Ne; but in effect two more species, $^{14}$N and $^{24}$Mg, are followed, by allowing for baryon conservation in the different burning regions. We also include the abundances of $^{28}$Si and $^{56}$Fe, but in the present code these elements are assumed to remain inert; however, they can be incorporated in an extended nuclear network. Our selection of five major species allows us to follow the evolution of stars to quite a late stage, in principle up to oxygen burning. The rates of 20 nuclear reactions, as listed in Table 1, are stored as tables for $6 < \log T < 10$.

The reaction rates are taken from Caughlan & Fowler (1988), except for the $^{12}$C$(\alpha, \gamma)^{16}$O reaction for which we use the rate from Caughlan et al. (1985), on W. Fowler's recommendation. Where two or more reactions are written in a chain, the later reactions are taken to be in transient equilibrium with the first, and their rates are not included.

**Table 1.** Nuclear reaction network.

---
$^1$H (p,$\beta^+\nu$) $^2$H (p,$\gamma$) $^3$He
$^3$He ($^3$He,2p) $^4$He
$^3$He ($^4$He,$\gamma$) $^7$Be
$^7$Be (e$^-$,$\nu$) $^7$Li (p,$\alpha$) $^4$He
$^7$Be (p,$\gamma$) $^8$B ($\beta^+\nu$) $^8$Be* ($\alpha$) $^4$He
$^{12}$C (p,$\beta^+\nu$) $^{13}$C (p,$\gamma$) $^{14}$N
$^{14}$N (p,$\beta^+\nu$) $^{15}$N (p,$\gamma$) $^{16}$O
$^{14}$N (p,$\beta^+\nu$) $^{15}$N (p,$\alpha$) $^{12}$C
$^{16}$O (p,$\beta^+\nu$) $^{17}$O (p,$\alpha$) $^{14}$N
$^4$He ($\alpha$) $^8$Be* ($\alpha$,$\gamma$) $^{12}$C
$^{12}$C ($\alpha$,$\gamma$) $^{16}$O
$^{14}$N ($\alpha$,$\gamma$) $^{18}$F ($\frac{1}{2}\alpha$,$\gamma$) $^{20}$Ne
$^{16}$O ($\alpha$,$\gamma$) $^{20}$Ne
$^{20}$Ne ($\alpha$,$\gamma$) $^{24}$Mg
$^{12}$C ($^{12}$C,$\alpha\gamma$) $^{20}$Ne
$^{12}$C ($^{12}$C,$\gamma$) $^{24}$Mg
$^{12}$C ($^{16}$O,$\alpha\gamma$) $^{24}$Mg
$^{16}$O ($^{16}$O,$\alpha\gamma$) $^{28}$Si ($\gamma$,$\alpha$) $^{24}$Mg
$^{20}$Ne ($\gamma$,$\alpha$) $^{16}$O
$^{24}$Mg ($\gamma$,$\alpha$) $^{20}$Ne
---

The first five reactions, comprising the pp-chain, are also assumed to be in transient equilibrium with each other, so that the (small) abundances of $^3$He and $^7$Be do not need to be followed explicitly, although they are in effect also followed. The reaction involving $^{18}$F is of course an invention, for convenience; in effect, $^{14}$N is taken to burn to $^{20}$Ne rather than to $^{22}$Ne, so that we can continue to use just the five nuclear species itemized above. Something similar applies to the last step in the (O,O) reaction. Other possible paths for the (C,C), (C,O) and (O,O) reactions are neglected. The enhancement of the reaction rates due to electron screening is taken into account with the prescription of Graboske et al. (1973).

## 4 RESULTS OF EVOLUTIONARY CALCULATIONS

To test the resulting code, we constructed a ZAMS from $\sim 0.08$ to $250\,\mathrm{M}_\odot$, and evolved 24 models of mass $0.64 - 125\,\mathrm{M}_\odot$, in steps of about 0.1 in the decimal log of mass, from the ZAMS to carbon ignition – except that the four lowest-mass stars left the asymptotic giant branch (AGB) before igniting carbon, and the four highest-mass stars broke down at an earlier stage, for reasons that are not yet clear. A subset of these models is shown in Fig. 4, along with the observational data that Andersen (1991) selected as the most reliable from eclipsing, double-lined spectroscopic binaries. We used a mixing-length ratio of 2.0, and zero-age composition $X = 0.7, Z = 0.02$. For the mixture of elements contained in $Z$ we adopted the ratios derived from Solar system meteorites by Anders & Grevesse (1989). At 'zero-age', we take the abundances to be uniform except that, in stars above $\sim 1.5\,\mathrm{M}_\odot$, $^{12}$C is assumed to have reached equilibrium through the CN cycle. The models were evolved at constant mass and with the Schwarzschild criterion to determine convective boundaries.

We were particularly concerned to test the following:

(a) whether reasonable models could be achieved near the bottom of the main sequence, where non-ideal effects most strongly affect the EOS;

(b) whether a model of $1\,\mathrm{M}_\odot$, at age 4.5 Gyr, would be located closer to the observed position of the Sun than we have found using the previous version of the code;

(c) whether models of $\sim 2-3\,\mathrm{M}_\odot$ would spend sufficient time in the blue subgiant region, above the 'hook' but not yet into thermal-time-scale evolution, to account for the surprisingly large number of observed systems (Andersen 1991) in this region;

(d) whether our first giant branch and AGB tracks for stars of $\sim 0.8 - 8\,\mathrm{M}_\odot$ would be significantly different from those of HPE (who used low-temperature molecular opacities from Weiss, Keady & Magee 1990), and whether this could in turn affect the relation found by HPE between initial (ZAMS) masses and the masses of white-dwarf remnants.

We hope to give a more detailed discussion of these points in a future paper. But, even in a detailed discussion, we can only make subjective judgments. Regarding point (a), we certainly were able to construct models down to $\sim 0.075\,\mathrm{M}_\odot$, where hydrogen burning gives out. The EOS did not produce obvious anomalies in such stars, although their interiors are very close to the region (Fig. 1) where difficult physics is to be encountered, and where our EOS is probably least reliable. Comparison with observation is very difficult, since the bolometric correction is very uncertain at these low temperatures.

Regarding point (b), our 1-$\mathrm{M}_\odot$ model passed very close to the position of the Sun (shown in Fig. 4), at an age of 5.3 Gyr. It is both cooler and fainter than the Sun at age 4.5 Gyr by 1 and 7 per cent respectively. With the EFF approximation, and older opacities, the discrepancies were much worse, about 5 and 20 per cent. However, the abundance ratios for the Sun as determined by Anders & Grevesse (1989) imply a value of $Z = 0.0188$ for an assumed $X = 0.7$. We constructed an opacity table for this metallicity (by linear interpolation in $Z$ between 0.01 and 0.02, the values for which OPAL data are available) and evolved a 1-$\mathrm{M}_\odot$ model with this metallicity and opacity table, assuming the same mixing length ratio as before. This model was still cooler and fainter than the Sun at 4.5 Gyr, but only by as little as 0.2 and 0.7 per cent. This is certainly within the accuracy that can be expected from either the code or the opacity tables, so that we did not attempt to 'improve' our solar model by adjusting the mixing length and the hydrogen abundance.

Regarding point (c), we note that, although at least six stars in Andersen's (1991) sample are, according to our models, evolved beyond the 'hook' at the end of core hydrogen burning, the length of time spent in the subgiant region, beyond the hook but on the left of the Hertzsprung gap, is not negligible. Fig. 4 attempts to show evolution on a nuclear time-scale as a solid line, and evolution on a thermal time-scale as a dotted line (with occasionally a dashed line showing an intermediate time-scale). The fractions of the main-sequence lifetime spent in the solid portion of the post-hook track, on the left of the Hertzsprung gap, are 5, 2 and 0.5 per cent respectively for 2, 4 and 8 $\mathrm{M}_\odot$. Although 0.5 per cent may seem very small, it is still more, by a factor of about 5, than the time spent on the whole of the subsequent dotted track. We believe that selection effects for bright, eclipsing binaries are quite likely to push up the representation of systems containing post-hook subgiants, perhaps by the necessary factor of three or four. Thus at $1.8 - 2.5\,M_\odot$, where Andersen's six stars lie, we are not convinced that there is a real discrepancy in frequency of these systems, especially since the statistical base is not large. Of course, a much more stringent test of our models would come from a comparison of individual binaries with evolutionary tracks, computed for the determined masses and (if available) metallicity. We hope to attempt such a study in the future.

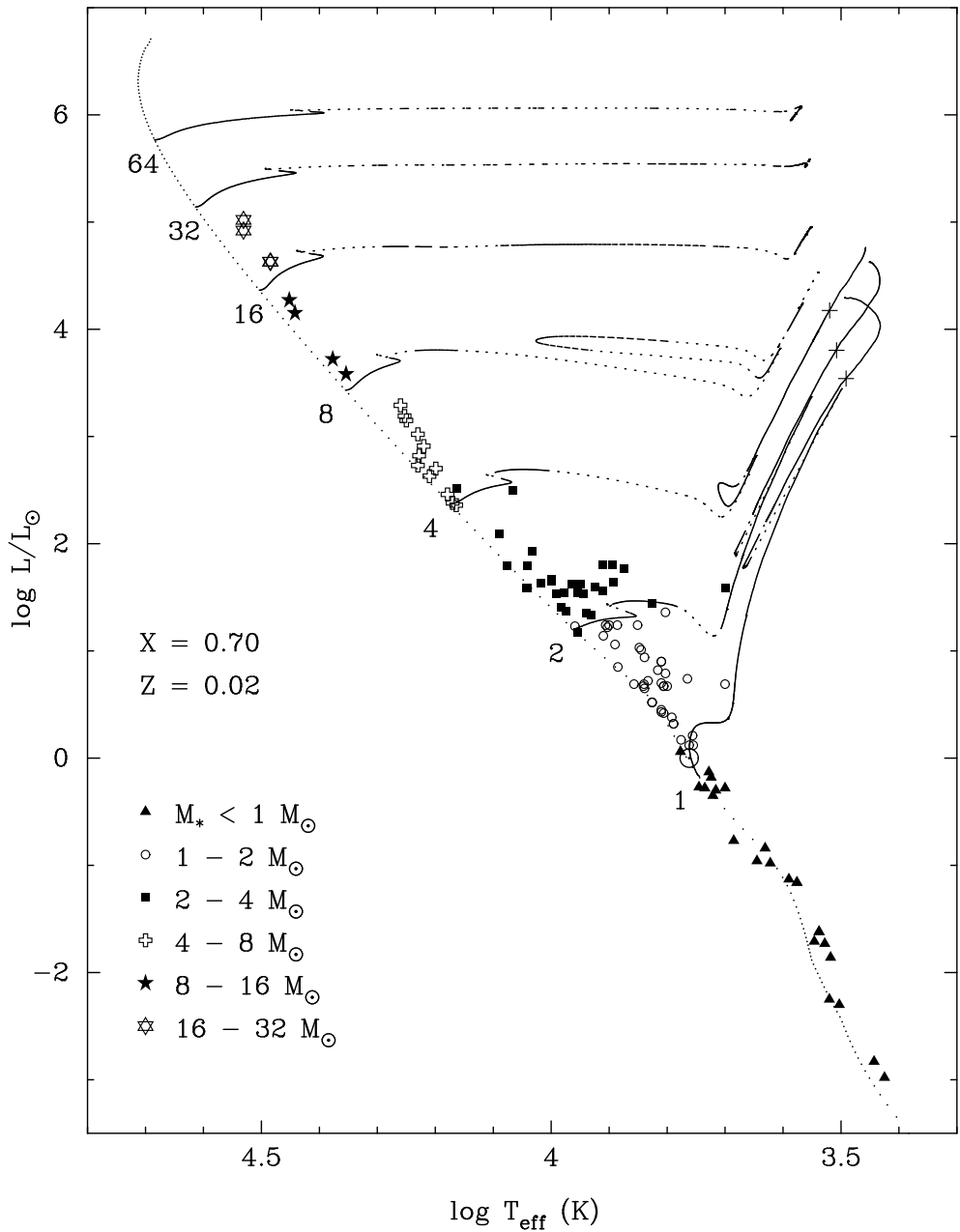

**Figure 4.** Hertzsprung–Russell diagram of the ZAMS and several evolutionary tracks, calculated with the updated stellar evolution code. The masses of the models are indicated at the starts of the tracks, in solar units. The solid portions of the tracks indicate where evolution is on a relatively slow nuclear time-scale, the dotted parts show evolution on a thermal time-scale, and the dashed parts show an intermediate time-scale. The different symbols indicate the positions of binary components with well-determined masses, radii and luminosities; data for eclipsing, double-lined spectroscopic binaries from Andersen (1991) and data for low-mass, visual binaries from Popper (1980). The position of the Sun is indicated by a solar symbol (⊙). The crosses (+) show the position on the giant branches of the 1-, 2- and 4-$M_\odot$ models where the binding energy of the envelope becomes positive (see text).

Regarding point (d), although our effective temperatures for highly evolved red giants are substantially higher than in HPE, mainly because the molecular opacities from Alexander & Ferguson (1994a, 1994b) are significantly different from those of Weiss et al. (1990), the points on the evolutionary tracks (shown by crosses) where the binding energy of the envelope goes through zero occur at very much the same luminosity and core mass. Thus we do not have to alter significantly the initial/final mass relation obtained by HPE. This relation (which includes no free parameters) was found by HPE to give a very good agreement with the distribution of masses of planetary-nebula nuclei (Jacoby 1989).

## ACKNOWLEDGMENTS

We would like to thank D. Alexander for his willingness to compute tables of molecular opacities on the grid requested by us. We also thank W. Däppen for helpful discussions, and W. Fowler and N. Itoh for useful suggestions. ORP and CAT are grateful for support by PPARC fellowships and research grants.

**Table A1.** Coefficients for the Fermi–Dirac integral expansion.

| $\hat{Q}_{mn}$ | | $m$ = 0 | 1 | 2 | 3 |
|---|---|---|---|---|---|
| | 0 | 1.157736 | 3.770676 | 4.015224 | 1.402284 |
| $n$ | 1 | 8.283420 | 26.184486 | 28.211372 | 10.310306 |
| | 2 | 14.755480 | 45.031658 | 46.909420 | 16.633242 |
| | 3 | 7.386560 | 22.159680 | 22.438048 | 7.664928 |

## APPENDIX A: MODIFICATION TO THE EVALUATION OF THE FERMI–DIRAC INTEGRALS

Since EFF, a slight variation has been introduced to the evaluation of one of the Fermi–Dirac integrals (cf. Webbink 1975). EFF gave an approximation to these integrals that has the correct asymptotic forms in all four limiting regions ($\psi \gg 1$, $\psi \ll -1$, $T \gg 1$ and $T \ll 1$) and is also a reasonable fit over the whole range of $\psi$ and $T$. The approximation is given in terms of $f$ and $g$ defined by

$$\psi = 2\sqrt{1+f} + \ln\frac{\sqrt{1+f}-1}{\sqrt{1+f}+1} \tag{A1}$$

and

$$g = \frac{kT}{m_e c^2}\sqrt{1+f}. \tag{A2}$$

In our revised equation of state we continue to use the approximations to $\rho^*$ and $P^*$ given by EFF (corresponding, within a numerical constant, to $n_e$ and $p_e$ respectively in the notation of this paper). We use their fourth-order expansion (table 5 of EFF) so that the error is about 0.3 per cent at worst. However, rather than calculating the internal energy $U^*$ directly, $Q^*$ was introduced to avoid problems with the cancellation of large numbers when finding the electron entropy. It is defined in a similar way to $\rho^*$ and $P^*$ by EFF:

$$Q^* \approx \hat{Q} = \frac{f}{(1+f)^5}\frac{g^{5/2}}{(1+g)^{3/2}}\sum_{m=0}^{3}\sum_{n=0}^{3}\hat{Q}_{mn}f^m g^n; \tag{A3}$$

the coefficients $\hat{Q}_{mn}$ are given in Table A1. As in EFF the coefficients were chosen so that the state functions are thermodynamically consistent; they obey Maxwell's relations exactly. The entropy per unit volume of the free electrons is then given by

$$s_e = n_e k s^*; \quad s^* = \sqrt{1+f}\left[\frac{Q^*}{g\rho^*} + 2\right] - \psi, \tag{A4}$$

and the internal energy by

$$\frac{U^*}{\rho^* T} = s^* + \psi - \frac{P^*}{\rho^* T}. \tag{A5}$$